\documentclass[a4paper]{article}
\usepackage{INTERSPEECH2019,multirow,subcaption,graphicx,comment}
\usepackage{ tipa }
\usepackage{enumitem}
\usepackage[T1]{fontenc}

\title{Neural Network-Based Modeling of Phonetic Durations}
\name{Xizi Wei\thanks{The first author is a PhD candidate at the University of Birmingham. The work was carried out during her internship at Apple UK.}, Melvyn Hunt, Adrian Skilling}
\address{Apple Inc} 
\email{\tt~ xxw395@bham.ac.uk, \{Melvyn\_Hunt,askilling\}@apple.com}
\begin{document}

\maketitle

\begin{abstract}
A deep neural network (DNN)-based model has been developed to predict non-parametric distributions of durations of phonemes in specified phonetic contexts and used to explore which factors influence durations most.
Major factors in US English are pre-pausal lengthening, lexical stress, and speaking rate. The model can be used to check that text-to-speech (TTS) training speech follows the script and words are pronounced as expected. Duration prediction is poorer with training speech for automatic speech recognition (ASR) because the training corpus typically consists of single utterances from many speakers and is often noisy or casually spoken. Low probability durations in ASR training material nevertheless mostly correspond to non-standard speech, with some having disfluencies. Children's speech is disproportionately present in these utterances, since children show much more variation in timing.  

\end{abstract}

\section{Introduction}

Much of the past work on phonetic duration falls into three categories, aimed at gaining phonetic insight, improving the quality of TTS and improving the accuracy of ASR. In the first category,  researchers have examined the extent to which  certain phonetic factors have an influence on duration (\textit{e.g.} lexical stress~\cite{Klatt1976LinguisticUO,stress1988}, pre-pausal lengthening~\cite{Klatt1976LinguisticUO,Campbell1991SegmentDI}, position~\cite{Luce1985ContextualEO}, word predictability~\cite{Brunelle2015EffectsOL,SherrZiarko2015WordFE,Bell2009} and speaking rate \cite{Chodrof2015}). 
Typically, only a single factor is studied at a time, and the amount of speech data is small and is taken from just one speaker or a small number of speakers (30 or fewer).  Some interesting linguistic questions have been investigated in this way~\cite{Brunelle2015EffectsOL,SherrZiarko2015WordFE,Bell2009}. In the second category, durations are modeled parametrically or non-parametrically using DNNs or LSTM-RNNs trained on much more data than in the first category to set the durations at runtime in a parametric speech synthesizer ~\cite{Henter2016RobustTD,DBLP:journals/corr/RonankiWKH16, Chen2017DiscreteDM,lstmduration}. Typically, hundreds of phonetic features are included and there is no attempt to study the influence of any of these features.
The third category is aimed at improving ASR accuracy by attempting to improve the weak duration modeling provided by standard HMM's using so-called Hidden Semi-Markov Models ~\cite{1168477,1659950}.  No insight is sought into the influences on duration in this category.  Although some improvements in accuracy have been claimed, the methods have not been widely adopted. This third category also includes duration modeling applied to speech recognition at the whole-word level~\cite{Ma2005ContextdependentWD,Power1996DurationalMF}, though this approach is effectively limited to small-vocabulary systems (specifically, digits), which are no longer widely used. 

The work reported here provides some insight into the phonetic factors controlling duration and aims ultimately to help improve both speech synthesis and recognition. A DNN is used to generate non-parametric output distributions over durations given the phonetic context for each phoneme.
We incorporate the duration factors in the model in three ways to investigate their effects on duration prediction individually or in group (see Section 3.2.1). 
From the output distributions given by the models with or without the lexical stress and pre-pausal information, we show that the DNN is able to learn the lengthening effect of these two features (Section 3.2.2). 
More data is used than in any other work we are aware of, both in speaker-specific investigations and in speaker-independent investigations, where data from tens of thousands of speakers is used.  
The most immediate application of this work is to training speech synthesis and recognition systems, where anomalous phonetic durations can indicate discrepancies between a transcription or script and what was actually spoken.

\vspace{-2mm}
\section{Method}

\subsection{Neural Network-Based Modeling}
\begin{figure}[ht]
    \centering
    \includegraphics[width=0.9\linewidth]{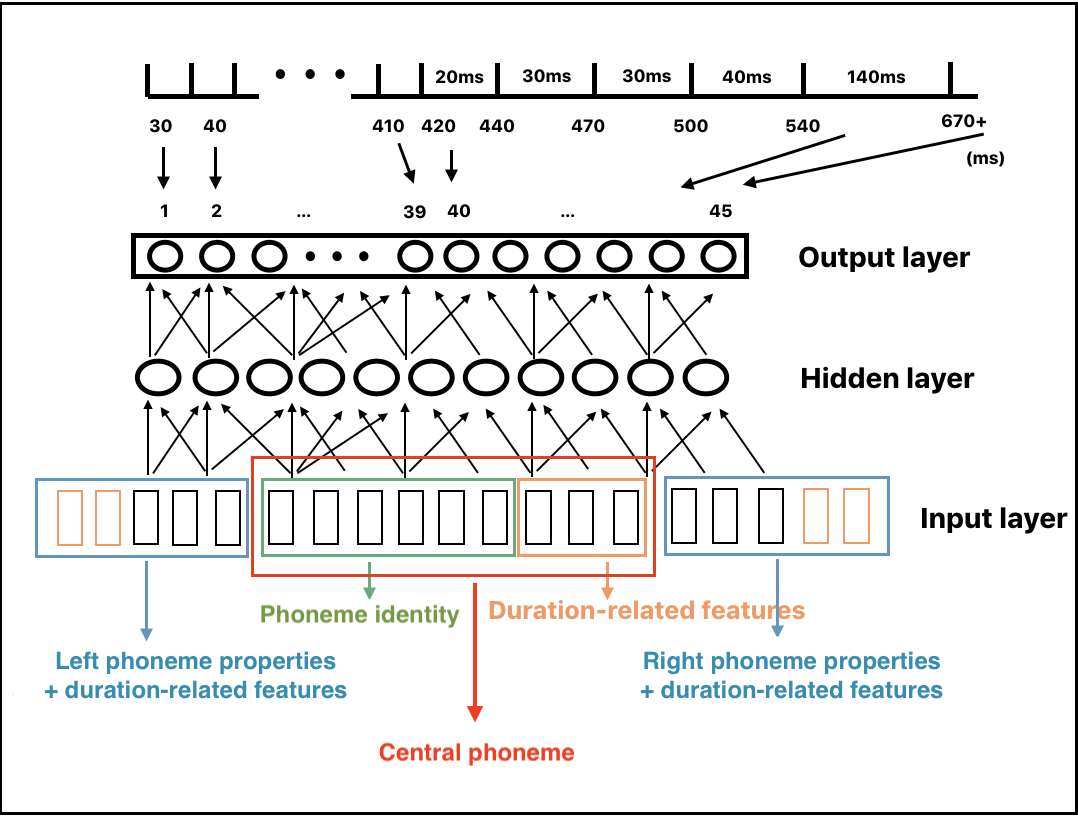}
    \caption{An overview of the duration modeling.
    }
    \label{fig:system overlook}
    \vspace{-2mm}
\end{figure}

\vspace{-2mm}
We used a feedforward DNN to model the duration, as shown in Figure~\ref{fig:system overlook}. The DNN comprises a stack of fully connected layers with the softmax function~\cite{John1990} at the output layer. We use the same number of units in each of the hidden layers. We use rectified linear activation (\textit{ReLU}) for the hidden units and cross-entropy as the loss function. The training procedure is optimized using \textit{ADAM}~\cite{DBLP:journals/corr/KingmaB14}. 
\vspace{-2mm}
\subsection{Input features}
\vspace{-2mm}
The inputs to the DNN are a concatenation of three types of information: identity of the current phoneme, phonetic properties of adjacent phonemes and duration-related features of the phonemes.
The identity of the current phoneme is encoded using a one-hot vector, while the phonetic properties of adjacent phonemes are characterized in a smaller vector (typically 15-dimensions). These phonetic properties include: long/short vowel, voiced/unvoiced consonant, plosive, affricate, nasal, fricative, glide, rhotic, sonorant, labial, alveolar, velar, aspirated and flap. The duration-related features are:

\begin{itemize}[noitemsep,leftmargin=*]
  \vspace{-1mm}
   \item \textbf{lexical stress}: 
   It has been widely reported that stressed syllables are usually longer than unstressed syllables~\cite{Klatt1976LinguisticUO,stress1988}. When stress information is available we use one bit to show whether the current phone is in a stressed syllable or not, and we also add the stress feature to the current phone and to adjacent phones, since the position relative to the stressed syllable also affects duration.
   \item \textbf{pre-pausal lengthening}: 
Speech sounds generally lengthen before a pause~\cite{Klatt1976LinguisticUO}. We add one feature to the central phone to indicate the distance between that phone and the next pause. The value = 1/n, when n, the number of phonemes to the following pause, =1,2,3,4 or 5; or 0 for n>5. 
  \item  \textbf{position in the syllable}:
   Draws information about position from this feature. One bit to show whether the current phone is a consonant preceding the vowel in a syllable. We add this feature to the side phones as well as to the current phone.

 \item \textbf{word predictability (LM scores)}: Studies in Vietnamese~\cite{Brunelle2015EffectsOL}, Mandarin~\cite{SherrZiarko2015WordFE} and English~\cite{Bell2009} have found that function words are spoken more quickly than non-function words and common words more quickly than rare words, suggesting that this behavior may be language-universal. These studies are consistent with the idea that words with a higher information content (\textit{i.e.} those that are less predictable) are spoken more carefully and hence more slowly. We use n-gram language model scores (on a log probability scale) to indicate the predictability of the word as an inverse measure of its information content.

 \item \textbf{speaking rate}: We use the ratio of the actual duration of the utterance to the duration the utterance would have given the expected phone durations as speaking rate. The expected phone duration is the average duration of that phone across the whole dataset.
 
\item \textbf{peak fundamental frequency (F0)}:
Peak F0 in the vowel was expected to influence duration through its association with \textit{focal lenthening}~\cite{focal_lengthening} in an utterance.  However, we were unable to find any influence of peak F0 on duration and it will not be discussed further.

\end{itemize}
\vspace{-4mm}
\subsection{Outputs}

We obtain the reference durations from forced-alignment.
We group them into 45 bins, starting at a bin corresponding to 30 ms, which is the shortest possible duration (one frame per state with three-state acoustic models) and increasing by 10 ms (one frame) for the first 39 bins. Beyond that point, there are too few samples at 10 ms spacing and the bins are made progressively wider.
For example, the 40th and 41st bins correspond to 20ms and 30ms spacing respectively. Durations larger than 670ms are all put into the 45th bin.

\vspace{-4mm}
\subsection{Outliers}
\vspace{-2mm}
We detect outliers using the output from the DNN. We use the value of the bin to which the reference duration belongs as the probability of the duration, and by ranking the probabilities we can get a list of phonemes with the lowest probabilities. These phonemes with unlikely duration, which we regard as outliers, can indicate misalignments or departures from the transcription. Examples are shown in Section 4.1.

\section{Experiments and Results}

\begin{table*}
\vspace{0mm} 
\caption{\it The duration prediction results for Baseline\_1 with different feature configurations trained on SPK1. The standard errors of the precisions (\%) are in the range from 0.0002 to 0.006. }
\vspace{-2mm} 
\centerline{
\begin{tabular}{|c|ccc|ccc|ccc|}
\hline
Models & \multicolumn{3}{c}{(i) Just the named feature}  & \multicolumn{3}{c}{(ii) Cumulative}  & \multicolumn{3}{c}{(iii) Leave one out} \vline\\  \cline{2-4} \cline{5-7} \cline{8-10}
 & precision & precision\_3 & CE\_loss   & precision & precision\_3 & CE\_loss & precision & precision\_3 & CE\_loss \\
pre/post-vocalic & 29.23 & 66.08 & 0.0296    & 29.23   &  66.08    & 0.0296  & 32.76 & 72.30 & 0.0273 \\ 
stress & 30.63& 68.47 & 0.0287 & 30.76 & 68.62 &0.0286 & 31.42 & 69.87 & 0.0282 \\
pre-pausal & 30.16 & 67.93& 0.0291& 32.28 & 71.26 & 0.0277& 31.81 & 70.49 & 0.0279\\
predictability & 29.19 & 66.08 & 0.0298& 32.52 & 71.70& 0.0276& 32.74 & 72.28 & 0.0272 \\
speaking rate & 29.54& 66.59 & 0.0294 & 33.08 & 72.73 & 0.0272&  32.52 & 71.70 & 0.0276 \\
\hline
\end{tabular}}
\vspace{-5mm}
\label{tabRes_features2}
\end{table*}

We measure the basic effectiveness of our modeling in three ways: (i) cross-entropy loss on a test set, (ii) a measure we call ``\textit{precision}'', which is the proportion of measured durations  whose bin exactly matches the mode of the model's predicted duration distribution. For most predictions (\textit{i.e.} those below 410ms), this is within 10 ms (one frame), and thus the highest precision possible, and (iii) a precision with more tolerance that counts not only the match to the bin corresponding to the peak of the distribution but also the neighboring bin on each side. We denote these three measurements as \textit{CE\_loss}, \textit{precision} and \textit{precision\_3} respectively.
All neural networks are built using \textit{Pytorch}~\cite{paszke2017automatic}.
\vspace{-2mm}
\subsection{Data}

We have in-house datasets from two native speakers of American English recorded for TTS purposes. One of the speakers SPK1, is female and the other, SPK2, male. There are 64,795 utterances (33 hours) in the SPK1 dataset and 27,550 utterances (13 hours) in the SPK2 dataset. We also have an in-house dataset, SPK-ASR, of less carefully controlled recordings intended for ASR that contains 540,389 utterances from 535,556 speakers of all ages, including children. We used forced alignment to get the duration for each phone. We used SPK1 and SPK2 for speaker-dependent modeling and SPK-ASR for speaker-independent modeling. The phonetic symbol sets used in these three datasets are different. We used 46, 42 and 50 one-hot vectors to encode the central phone identity for the three datasets respectively.
\subsection{Speaker-dependent modeling}
\subsubsection{Duration-related features configurations}
We used a DNN with 2 hidden layers and 256 hidden units in each layer as a baseline for exploring the feature configurations. 
We used a minibatch of 64 and train the model for 30 epochs. The learning rate was 0.001 for each epoch. 
We found that the final result depended to some extent on the random start point of the model built.  We therefore ran our precision tests 10 times, each with a different random start and a different randomly selected test set. 
For each test, we randomly sampled the dataset to use 90\% for training and 10\% for testing. 
We then computed the overall mean and standard error of the precision.
We began with \textit{Baseline\_0} trained on SPK1 (the input to the neural network being just the one-hot vector that encodes the identity of the current phone with no context) and obtained a precision of 19\%. In \textit{Baseline\_1} the input has a context of $\pm 1$ (1 phone on each side of the current phone) and the precision increased to 28.68\%. Thus, we obtained a 9.28\% absolute increase by including context.

\begin{figure}[ht]
    \centering
    \includegraphics[width=1\linewidth]{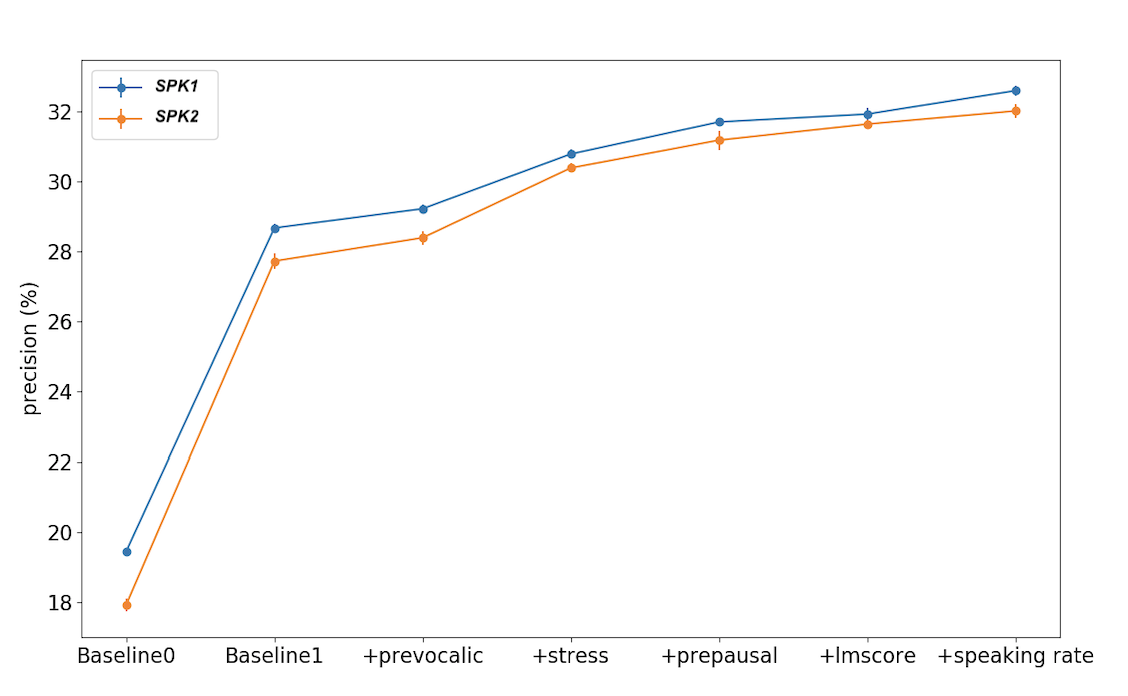}
    \vspace{-4mm}
    \caption{Models with different feature configurations trained separately on two speakers: SPK1 and SPK2. 
    }
      \vspace{-7mm}
    \label{fig:feature}
  
\end{figure}

Baseline\_1 was then augmented with the duration-related features in three ways: (i) adding each one to the Baseline\_1 to see the effect on its own, (ii) cumulatively adding the features to the Baseline\_1 and (iii) including all the features except the named one. The results are shown in Table~\ref{tabRes_features2}. 
The precision increases as the duration-related features are added, among which the stress has the biggest positive effect. The speaking rate for the utterance and pre-pausal lengthening also have a strong influence. The location of consonants within a syllable (\textit{i.e.} before or after the vowel) has a somewhat weaker influence, as has the predictability of the word containing the phoneme as estimated by a stochastic language model. We also carried out the cumulative experiments on another TTS dataset, SPK2. Figure~\ref{fig:feature} shows that the effect of these factors is very similar for a different speaker.
 
\subsubsection{Stress and pre-pausal lengthening effect}
\begin{figure}[ht]
    \centering
    \begin{subfigure}[b]{0.45\textwidth}
    \includegraphics[width=1\linewidth]{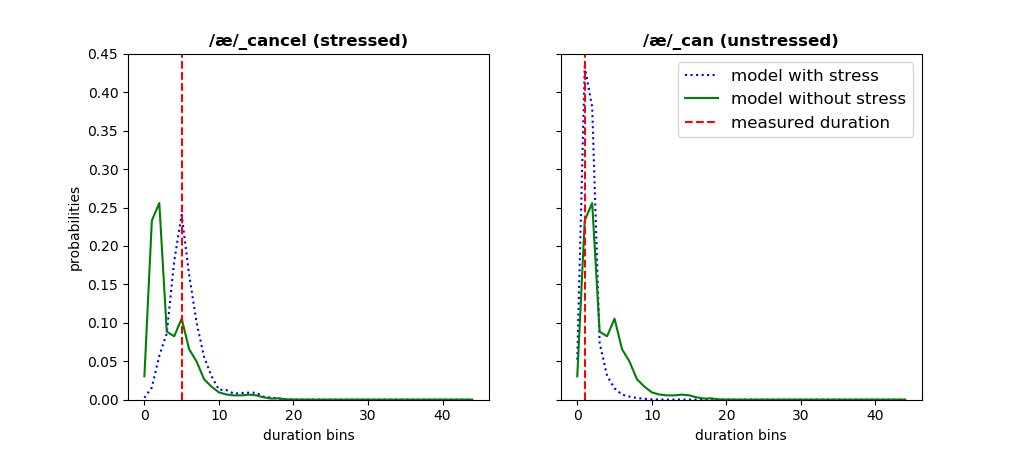}
    \caption{/\textnormal{{\ae}}/  in ``cancel'' and ``can''  comparing model outputs when lexical stress information is or is not included. }
    \label{fig:lengthening}
    \end{subfigure}
    \begin{subfigure}[b]{0.45\textwidth}
    \includegraphics[width=1\linewidth]{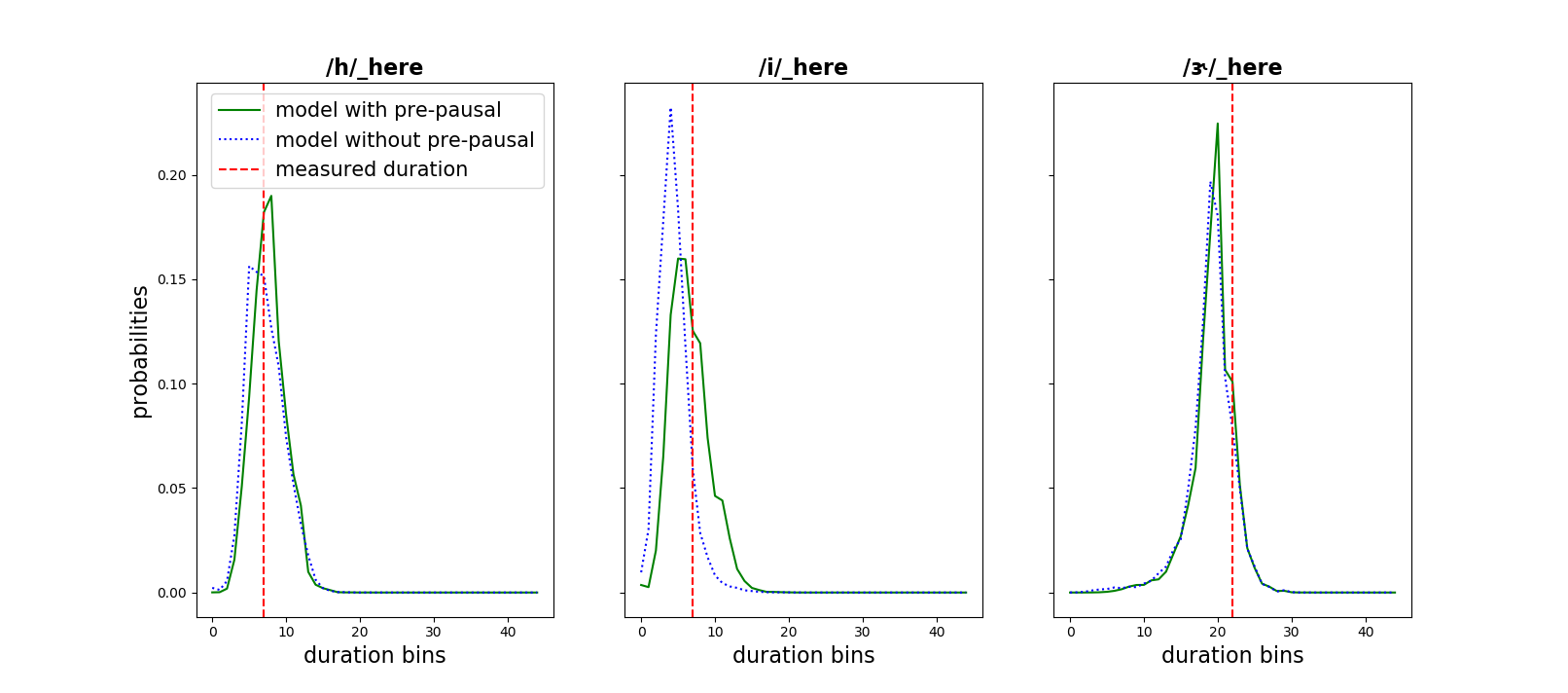}
    \caption{/h/ /i/ /\textrhookrevepsilon/ in an utterance-final ``here'', comparing distributions with and without an input providing the distance to the next pause.}
    \label{fig:lengthening2}
    \end{subfigure}
\caption{ Predicted duration distributions. The red dotted line shows the measured duration for one example.}
\label{fig:lengthing3}
\end{figure}

 \vspace{-2mm}
The duration probability distributions in Figure~\ref{fig:lengthing3} give examples showing that the network is able to learn the lengthening effect of stress and pre-pausal features, and the predictions are closer to the measured duration bins (red dashed lines) with these two features on. The /{\ae}/ in ``cancel'' and ``can'', which we denote as ``{\ae}\_cancel'' and ``{\ae}\_can'', have the same context, but different stress values (``can'' as a modal verb normally being unstressed).
In Figure~\ref{fig:lengthening}, the two green curves are the same, but knowledge of stress increases the predicted duration for /{\ae}/ in ``cancel'' and reduces it in ``can''.
Figure~\ref{fig:lengthening2} compares distributions with and without an input providing the distance to the next pause.  
Predicted duration distributions for utterance-final ``here'' are shown left to right as the three phonemes /h/ /i/ /\textrhookrevepsilon/.
Since the baseline model here has a context of $\pm 1$, the distribution of /\textrhookrevepsilon/ in ``here'' still has the effect of pre-pausal lengthening even without that feature given in the input features. For the /h/ and /i/, knowledge of pause proximity increases the predicted duration. 

We evaluated the SPK2-trained model on the SPK1 testing set and the SPK1-trained model on the SPK2 test set. Since SPK1 has much more training data than SPK2, we also evaluated the SPK1-trained model with reduced training data size. The results shown in Table~\ref{tab:corss_speaker} suggest that the precision decreases considerably when testing on a different speaker.  The results in row 3 with a model trained on SPK1 when using a reduced set to match that available for SPK2 match much more closely the results from training on SPK2 (row 1), suggesting that the difference between the results with the two speakers is largely attributable to the discrepancy in the amount of training material and indicating that more than 10 hours of training speech is needed for optimal model training.  This result also largely explains the offset between the two curves in Figure~\ref{tab:corss_speaker}.

\begin{table}[ht]
    \centering
    \caption{Cross-speaker precision tests (\%). The models used all the duration-related features.}
    \begin{tabular}{|ccc|}
    \hline
        training &  SPK2\_test (1h) & SPK1\_test (3h) \\
    \hline
         SPK2 (10h) & 31.00 & 22.70\\
         SPK1 (30h)& 23.34 & 32.56 \\
         SPK1 (10h)& 22.84 & 31.45 \\
    \hline
    \end{tabular}
    \vspace{2mm}
    \label{tab:corss_speaker}
    \vspace{-8mm}
\end{table}

\subsubsection{Model configurations}
  \vspace{-2mm}
We trained the DNN with a range of hidden layers ($d \in 1,2,3 $)
and hidden units in each layer ($w \in 128,256,512$)
and wider context ($\pm 1$,  $\pm 2$ and  $\pm 3$). 
\begin{figure}[ht]
    \centering
    \includegraphics[width=1\linewidth]{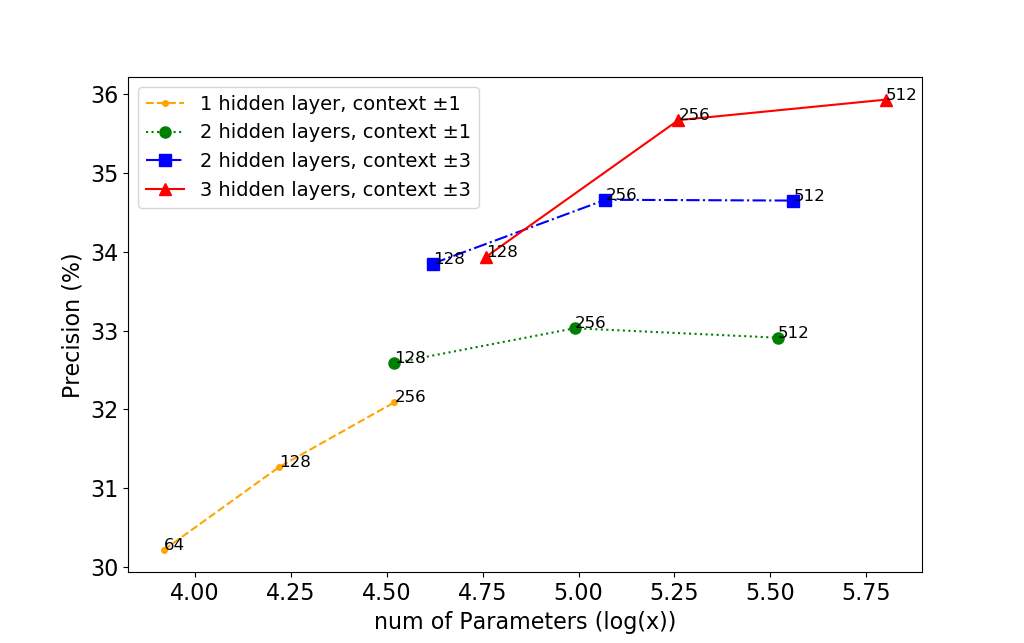}
    \caption{Model configurations.
    }
    \label{fig:system conf}
    \vspace{-2mm}
\end{figure}
As shown in Figure~\ref{fig:system conf}, the precision improves as the number of parameters in the model is increased and the context is longer. Taking computational efficiency into account, the best configuration for now is three hidden layers with 256 hidden units in each layer and with $\pm 3$ context, which achieves a precision of 35.67\% and precision\_3 of 89.88\%.

  \vspace{-2mm}

\subsection{Speaker-independent modeling}
  \vspace{-2mm}
We applied our duration modeling method with the configuration in Section 3.2.3 to speaker-independent modeling with 80\% of the SPK-ASR dataset as the training data. We obtained a precision of 10.50\% and precision\_3 of 40.30\% on a testing set (10\%). It is more challenging because in the SPK-ASR corpus almost every utterance is from a different speaker and spoken in a spontaneous way. Moreover, stress and LM score have not yet been incorporated.

\section{Applications}
\subsection{Outlier detection in TTS and ASR}

We use the best configuration from Section 3.2.3 to detect outliers for the SPK1 TTS dataset and the ASR dataset.

\begin{figure}[ht]
    \centering
    \begin{subfigure}[b]{0.45\textwidth}
    \includegraphics[width=1\linewidth]{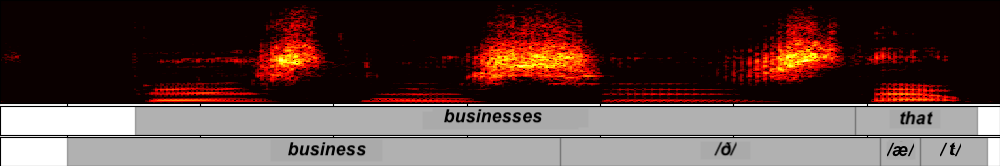}
    \caption{deviation from the script: the speaker says ``businesses'', but the transcription has ``business''; the /\dh/ in ``that'' is consequently misaligned to the end of ``businesses''. }
    \label{fig:tts1}
    \end{subfigure}
    \begin{subfigure}[b]{0.45\textwidth}
    \includegraphics[width=1\linewidth]{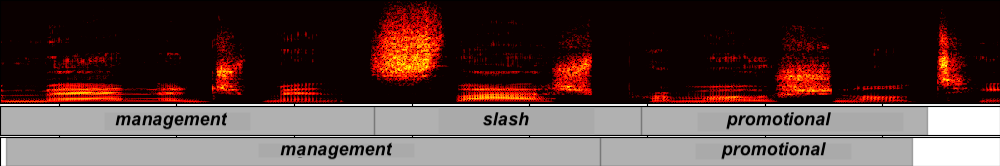}
    \caption{deviation from the script: the speaker says ``management slash promotional'', having evidently read ``management/promotional'', but the transcription has ``management promotional'', thus the /t/ is misaligned to an unlikely long duration.}
    \label{fig:tts2}
    \end{subfigure}
    \begin{subfigure}[b]{0.45\textwidth}
    \includegraphics[width=1\linewidth]{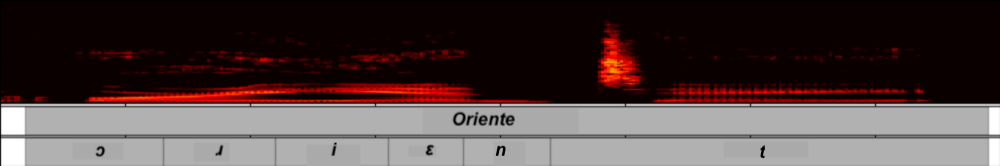}
    \caption{mismatch in the way the word is pronounced relative to the dictionary: the speaker says ``Oriente'' as /\textopeno\textturnr i\textquotesingle\textepsilon nte/, but the pronunciation in the dictionary for ``Oriente'' is /a\textturnr i\textquotesingle\textepsilon nt/ without a final vowel,  causing the /t/ to be misaligned to a longer segment.}
    \label{fig:tts3}
    \end{subfigure}
    \begin{subfigure}[b]{0.45\textwidth}
    \includegraphics[width=1\linewidth]{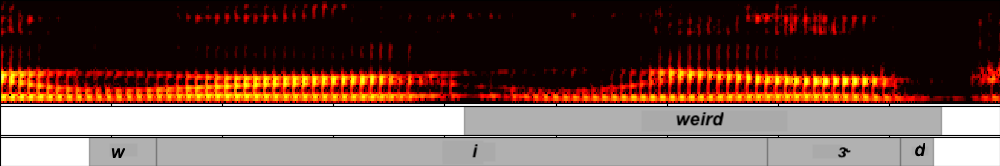}
    \caption{mistranscription: the speaker actually says ``wei... weird'' but the transcription is simply ``weird''; the /i/ is consequently aligned to a much longer portion of speech.}
    \label{fig:asr1}
    \end{subfigure}
\caption{Outlier examples, the upper annotation line is what the speaker says and the lower is from the forced alignment.}
  \vspace{-4mm}
\end{figure}

Figure~\ref{fig:tts1},~\ref{fig:tts2} and~\ref{fig:tts3} show three examples from the top outliers in the SPK1 dataset corresponding to three kinds of problems that have been seen to occur in the TTS training corpus. 
48 out of 50 outliers are correctly detected as having bad alignments.

We also observed misalignments in the ASR dataset as well as disfluencies as in Figure~\ref{fig:asr1} (such disfluencies being rare in the speech of the professional speakers producing the TTS dataset). 
When examining outliers in ASR (Table~\ref{tab:outliers}), 12 out of the top 50 outliers (\textit{i.e.} 24\%) were found to be from children.  By contrast, just 11 out 100 randomly selected utterances were judged to be from children, suggesting that children make a disproportionate contribution to the set of outliers. Among the top 50 outliers, 8\% are because of disfluencies, resulting in bad transcriptions and hence bad alignments. 
 By contrast, in the randomly selected utterances, fewer than 2\% were found to have bad alignments.
 The rest of the outliers evidently get their low scores because the speaker was dictating and hence speaking slowly, had put extreme stress on a word and hence lengthened it, or was speaking in a playful style. 

\begin{table}[ht]
    \centering
    \caption{Proportion of children's speech and bad alignments in the top 50 outliers and the randomly selected utterances for the ASR data.}
    \begin{tabular}{|ccc|}
    \hline
       &  Top 50 outliers & Random utts \\
    \hline
         Children's speech & 24\% & 11\%\\
         Bad alignments & 8\% & <2\% \\
    \hline
    \end{tabular}
    \vspace{2mm}
    \label{tab:outliers}
\end{table}
\vspace{-8mm}

\section{Conclusions}

A DNN can provide a useful prediction of the distribution of durations of a phoneme in a specified context.  It offers a technique for gaining a basic understanding from large speech corpora (rather than the more usual small set of examples) of how various factors combine to determine phonetic durations in a given language.
The prediction is best, at least in American English, when the phonetic properties of at least three phonemes on each side of the phoneme under consideration are provided to the DNN, together with other relevant information, such as lexical stress in the syllable and estimated average speaking rate.

Distributions produced in this way can be used to spot improbable durations that often arise from a mismatch between the speech and either the words in the phonetic transcription or the dictionary pronunciations of those words.
Low probability durations may also occur because the speech is particularly expressive. In training material for TTS these anomalies can be used to correct transcriptions and dictionary entries as well as to exclude unsuitable speech from the TTS training set. In ASR training material, low duration scores may result from disfluencies (rare in TTS training speech), but the most common cause from our limited sampling of the outliers appears to be unusual timing from expressive speech or dictation mode.  This second cause does not invalidate the speech for ASR training purposes, though the first clearly does.  Children's speech is overrepresented in the set of low duration scores because phonetic durations in their speech appear to be much more variable than those of adults' speech.

So far, this work has been confined to American English.  We might speculate that duration information will be particularly useful for ASR in languages such as Japanese, Finnish, Estonian and Arabic~\cite{Zangar2018} that have phonemic length.

\section{Acknowledgments}
Earlier work on seeking factors that influence durations was carried out by Dominic Hunt and especially by Paul Coles.  John Bridle, Barry Theobald and Rin Metcalf made many useful comments and Felipe Espic pointed us to the \textit{ADAM} optimizer.
\bibliographystyle{IEEEbib}
\bibliography{reference}

\end{document}